\documentclass[twocolumn]{aastex631}

\hypersetup{linkcolor=red,citecolor=blue,filecolor=cyan,urlcolor=cyan}
\usepackage{graphicx}
\usepackage{amsmath}
\usepackage{amsbsy}
\usepackage{soul}
\shorttitle{Obliquity of TOI-677\,b}
\shortauthors{Sedaghati et al.}
\graphicspath{{./}{figures/}}
\begin{document}

\title{Orbital alignment of the eccentric warm Jupiter TOI-677\,b}

\correspondingauthor{Elyar Sedaghati}
\email{esedagha@eso.org}

\author[0000-0002-7444-5315]{Elyar Sedaghati}
\affiliation{Facultad de Ingenier\'ia y Ciencias, Universidad Adolfo Ib\'a\~nez, Av.\ Diagonal las Torres 2640,
Pe\~nalol\'en, Santiago, Chile}
\affiliation{Millennium Institute for Astrophysics, Chile}
\affiliation{European Southern Observatory (ESO), Av. Alonso de Córdova 3107, 763 0355 Vitacura, Santiago, Chile}

\author[0000-0002-5389-3944]{Andr\'es Jord\'an}
\affiliation{Facultad de Ingenier\'ia y Ciencias, Universidad Adolfo Ib\'a\~nez, Av.\ Diagonal las Torres 2640, Pe\~nalol\'en, Santiago, Chile}
\affiliation{Millennium Institute for Astrophysics, Chile}
\affiliation{Data Observatory Foundation, Chile}

\author[0000-0002-9158-7315]{Rafael Brahm}
\affiliation{Facultad de Ingenier\'ia y Ciencias, Universidad Adolfo Ib\'a\~nez, Av.\ Diagonal las Torres 2640,
Pe\~nalol\'en, Santiago, Chile}
\affiliation{Millennium Institute for Astrophysics, Chile}
\affiliation{Data Observatory Foundation, Chile}

\author[0000-0003-2186-234X]{Diego J. Muñoz}
\affiliation{Facultad de Ingenier\'ia y Ciencias, Universidad Adolfo Ib\'a\~nez, Av.\ Diagonal las Torres 2640, Pe\~nalol\'en, Santiago, Chile}
\affiliation{Millennium Institute for Astrophysics, Chile}
\affiliation{Data Observatory Foundation, Chile}
\affil{Center for Interdisciplinary Exploration and Research in Astrophysics (CIERA) and Department of Physics and Astronomy, Northwestern University, 2145 Sheridan Road, Evanston, IL 60208, USA}

\author[0000-0003-0412-9314]{Cristobal Petrovich}
\affiliation{Millennium Institute for Astrophysics, Chile}
\affiliation{Instituto de Astrofísica, Pontificia Universidad Católica de Chile, Av. Vicuña Mackenna 4860, 782-0436 Macul, Santiago, Chile}

%\author[0000-0002-3728-3329]{Pablo Benítez-Llambay}
%\affiliation{Facultad de Ingenier\'ia y Ciencias, Universidad Adolfo Ib\'a\~nez, Av.\ Diagonal las Torres 2640, Pe\~nalol\'en, Santiago, Chile}

\author[0000-0002-5945-7975]{Melissa J. Hobson}
\affiliation{Millennium Institute for Astrophysics, Chile}
\affiliation{Max-Planck-Institut für Astronomie, Königstuhl 17, D-69117 Heidelberg, Germany}

%\collaboration{6}{(AAS Journals Data Editors)}



\begin{abstract}
Warm Jupiters lay out an excellent laboratory for testing models of planet formation and migration. Their separation from the host star makes tidal reprocessing of their orbits ineffective, which preserves the orbital architectures that result from the planet-forming process. Among the measurable properties, the orbital inclination with respect to the stellar rotational axis, stands out as a crucial diagnostic for understanding the migration mechanisms behind the origin of close-in planets. Observational limitations have made the procurement of spin-orbit measurements heavily biased toward hot Jupiter systems. In recent years, however, high-precision spectroscopy has begun to provide obliquity measurements for planets well into the warm Jupiter regime. In this study, we present Rossiter-McLaughlin (RM) measurements of the projected obliquity angle for the warm Jupiter TOI-677\,b  using ESPRESSO at the VLT. TOI-677\,b exhibits an extreme degree of alignment ($\lambda$\,$=$\,$0.3$\,$\pm$\,$1.3$\,deg), which is particularly puzzling given its significant eccentricity ($e$\,$\approx$\,0.45). TOI-677\,b thus joins a growing class of close-in giants that exhibit large eccentricities and low spin-orbit angles, which is a configuration not predicted by existing models. We also present the detection of a candidate outer brown dwarf companion on an eccentric, wide orbit ($e$\,$\approx$\,0.4 and $P$\,$\approx$\,13\,yr). Using simple estimates, we show that this companion is unlikely to be the cause of the unusual orbit of TOI-677\,b. Therefore, it is essential that future efforts prioritize the acquisition of RM measurements for warm Jupiters.
\end{abstract}



\section{Introduction} \label{sec:intro}
Warm giant planets, those with radii comparable to that of Jupiter and orbital periods in the range of $\sim$\,10\,$-$\,200\,days, are well suited for advancing our understanding of close-in giant planet formation. In contrast to their hotter counterparts --the so-called hot Jupiters (period $\lesssim$\,10\,d)-- warm giants are not expected to be subject to significant tidal friction \citep[e.g.,][]{Alexander1973,Zahn1977,Hut1981}, thus better preserving their primordial orbital configurations. Consequently, 
characterisation of warm giant orbits, albeit a significant observational challenge, can help better constrain
 planet formation models. 
 
Mechanisms through which close-in giant planets form are hotly debated, but generally speaking, there are two families of models: (i) {\it in situ} formation and (ii) planetary migration. In situ scenarios rely on the core accretion model \citep{Pollack1996} to work at small stello-centric distances, provided there is enough gas, and that critical cores can form from a sufficiently dense distribution of solids \citep[e.g.,][]{Batygin2016} or from the consolidation of smaller cores \citep{Boley2016}. Planetary migration, on the other hand, relies on the significant reduction of planet's semi-major axis from initial separations beyond the ice line. Migration can be mediated by the tidal interaction with a gaseous, Keplerian disk \citep{Lin1979,Goldreich1980,Ward1997} or mediated by extreme eccentricity growth followed by circularization and orbital decay \citep[e.g.,][]{Mazeh1979}, which result naturally from tidal friction \citep[e.g.,][]{Goldreich1963,Goldreich1966,Hut1981}. This ``high-eccentricity migration'' can be triggered by planet-planet scattering  \citep[e.g.][]{Rasio1996} or by different types of secular perturbations \citep[e.g.,][]{Eggleton2001,Wu2003,Fabrycky2007,Wu2011,Naoz2011,Petrovich2015}.

One may also choose to categorize these different formation mechanisms as either ``dynamically cold'' or ``dynamically hot'' \citep[e.g.,][]{Tremaine2015}. In dynamically cold channels,
the eccentricities, inclinations and obliquities remain low; in dynamically hot evolution, on the other hand, the orbital elements can vary widely. For instance, in situ formation and disk-driven migration do not typically involve growth in inclination nor eccentricity, and can be deemed dynamically cold. High-eccentricity migration, on the other hand, is by definition, dynamically hot. Thus, measuring a warm giant's eccentricity and/or inclination relative to the stellar spin axis can serve as a discriminant between ``hot'' and ``cold'' dynamical histories, and consequently, serve as a crucial diagnostic of planet formation theories.

In principle, a sufficiently large number of spin-orbit measurements could prove extremely powerful for discerning between different planet migration models \citep[e.g.,][]{Morton2011}. Nonetheless, measurement of the spin-orbit angle (or projected stellar obliquity) $\lambda$ is more difficult for warm Jupiters than for hot Jupiters, due to the rarity and longer duration of their transits. The angle between the stellar rotation axis and the planet’s angular momentum vector, projected onto the plane of the sky, is measured through the observations of the Rossiter-McLaughlin (RM) effect \citep{Rossiter1924,McLaughlin1924} with spectroscopic observations during the exoplanet transit, which has thus far limited these observations to close-in planets around bright stars. Recently, however, high resolution spectroscopic observations at large aperture telescopes have made RM measurements of warm Jupiter systems possible, suggesting that these planets might represent a population significantly different from their hotter counterparts \citep{Rice2022}.

In this work we present the sky-projected obliquity measurement for the warm Jupiter planet TOI-677\,b \citep{Jordan2020}, through the analysis of the RM effect, observed with high resolution, spectroscopic observations of a single primary transit of the exoplanet.  This study is structured as follows: in \S~\ref{sec:intro} an introduction to the analysis is presented; in \S~\ref{sec:obs} we briefly present the observations of the target with ESPRESSO and the subsequent data reduction process; in \S~\ref{sec:analysis} the underlying analytical model, as well as the non-parametric noise model are presented, as well as the determination of the orbital obliquity angle from the modeling of the ESPRESSO transit data, while additional FEROS radial velocity observations are analysed together with previous data to infer a possible presence of an outer companion in the system; in \S~\ref{sec:discussion} we discuss the possible implications of our results in the greater context of giant-planet formation theories; and finally in \S~\ref{sec:conclusions} we summarise this work and present the final conclusions of the study. 

\section{Observations \& data reduction} \label{sec:obs}
\begin{figure}[t]
    \centering
    \includegraphics[width=\linewidth]{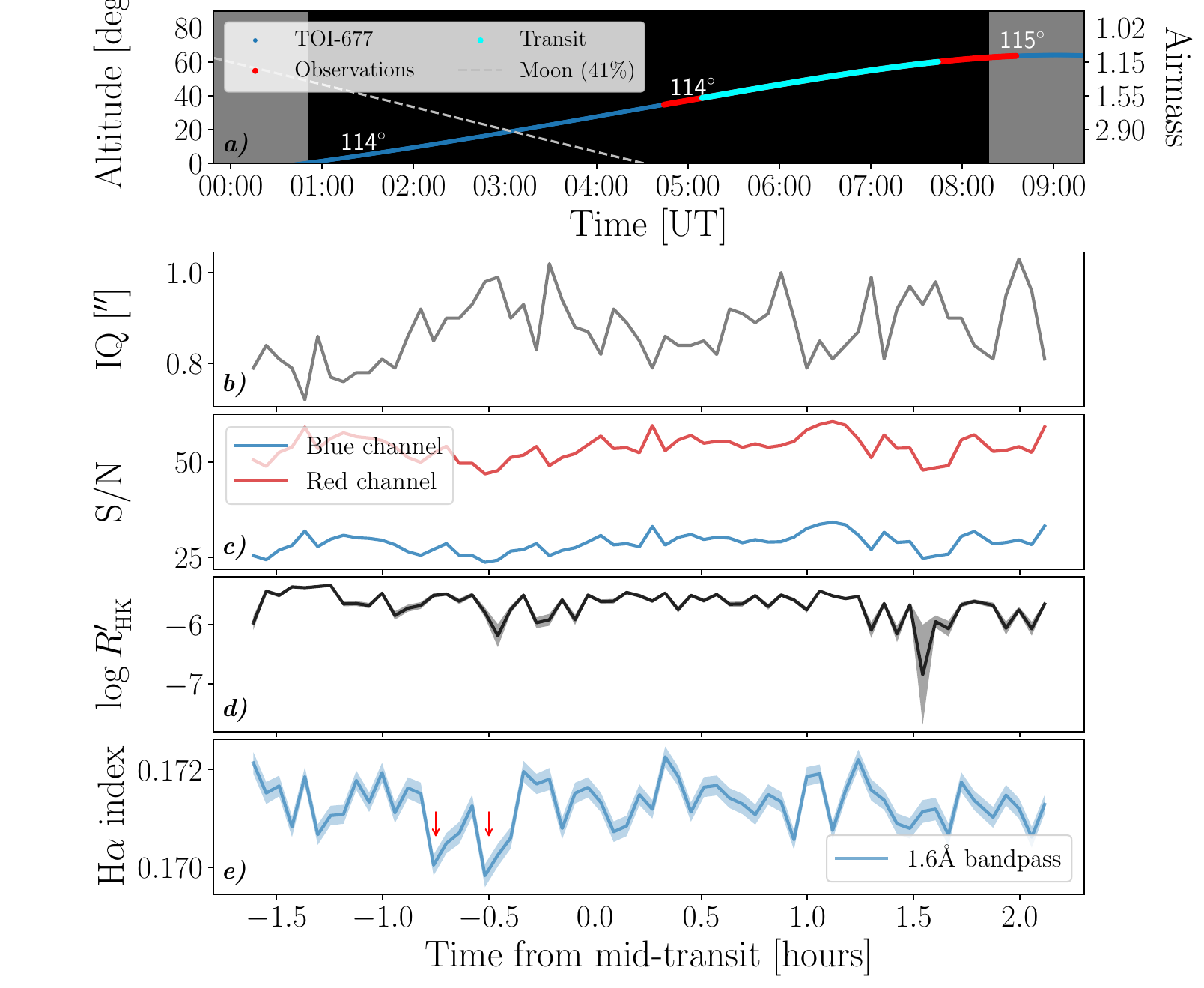}
    \caption{\textit{a)} altitude of the target star TOI-677 during the night of observations, with the duration of ESPRESSO exposures highlighted in red and the full transit in cyan. The numbers in the plot represent the angular separation of the target and the moon at each instance. \textit{b)} the variations of image quality delivered at the detector and corrected for the airmass. \textit{c)} the median spectral S/N values in the echelle orders recorded by the blue and red detectors. \textit{d)} variations of the $\log R^{\prime}_{\textrm{HK}}$ activity index, indicative of the chromospheric contributions of the H and K Calcium lines, excluding the photospheric component. The shaded regions present the 1$\sigma$ uncertainty, estimated by the ESPRESSO pipeline. \textit{e)} the variations of the H$\alpha$ index, calculated for the wider 1.6\,\AA~central bandpass, with the shaded region again representing the 1$\sigma$ uncertainty \citep{Gomes2011}. The two red arrows indicate the times at which variation in the index coincides with systematic noise in the RVs, resulting in deviations from the expected RM effect (c.f. Figure \ref{fig:final fit}).}
    \label{fig:observations}
\end{figure}
We observed a single primary transit of TOI-677\,b on the 9th of December 2021, with the ESPRESSO spectrograph \citep[Echelle SPectrograph for Rocky Exoplanets and Stable Spectroscopic Observations;][]{Pepe2021}, installed at the Incoherent Combined Coudé Focus (ICCF) of ESO's Very Large Telescope (VLT) at Paranal Observatory, Chile. TOI-677\,b is a 1.24\,$\pm$\,0.07\,M$_\textrm{J}$, 1.17\,$\pm$\,0.03\,R$_\textrm{J}$ planet on an eccentric ($e=0.44\pm0.02$) 11.2366\,$\pm$\,0.0001\,day orbit around a late F-type star. This host star has an effective temperature of 6295\,$\pm$\,77\,K, with $\nu\sin{i}$\,$=$\,7.80\,$\pm$\,0.19\,km/s \citep{Jordan2020}, estimated with the {\tt zaspe} code \citep{Brahm2017}. The stellar parameters determined in the detection study have been summarized in Table \ref{tab:stellar pars}, in addition to some of those that have been determined from the spectral synthesis analysis of the out of transit ESPRESSO spectra obtained in this study, using {\tt zaspe}.

\begin{deluxetable}{lcc}[h!]
\tabletypesize{\small}
\tablewidth{0pt} 
%\tablenum{1}
\tablecaption{Stellar parameters of TOI-677. \label{tab:stellar pars}}
\tablehead{
\colhead{Parameter} & \colhead{\citet{Jordan2020}} & \colhead{This work} }
%\colnumbers
\startdata 
Age [Gyr] & $2.92^{+0.80}_{-0.73}$ & $3.1 \pm 0.7$ \\
J-band magnitude, $m_J$  & $8.722 \pm 0.020$ & -- \\
Mass, M$_\star$ [M$_\odot$] & $1.17 \pm 0.06$ & $1.158^{+0.029}_{-0.027}$\\
Radius, R$_\star$ [R$_\odot$] & $1.28 \pm 0.03$ & $1.281 \pm 0.012$\\
Temperature, T$_{\textrm{eff}}$ [K] & $6295 \pm 77 $ & $6295 \pm 80$\\
$\log g$\,[dex] & $4.291 \pm 0.025$ & $4.286^{+0.016}_{-0.015}$\\
Metallicity, [Fe/H]\,[dex] & $0.00 \pm 0.05$ & $-0.02 \pm 0.05$\\
$\nu \sin{i_\star}$\,[km/s] & $7.80 \pm 0.19$ & $7.42 \pm 0.5$\\
\enddata
\end{deluxetable}

The observations were performed in the single-UT, HR mode (i.e.\ using the 1$^{\prime\prime}$ entrance fiber) with Unit Telescope 1 (UT1). The spectrograph records cross-dispersed echelle spectra through two blue- and red-optimised cameras, at a median resolving power $\mathcal{R}$\,$\approx$\,140\,000. The detectors were read in the unbinned readout mode at an average spectral sampling of 4.5\,pixels per resolution element, where each spectral order is recorded onto two slices owing to the anamorphic pupil slicing unit (APSU) of the spectrograph. Starting at 04:49\,UT, a total of 62 spectra were recorded (6 before, 40 during and 16 after transit) at exposure times of 180\,s, with S/N values of $\approx$\,70 at 550\,nm, across the two slices. A more detailed view of the observations is presented in the top panel of Figure \ref{fig:observations}. The observations were performed with the principal fibre (A) on the target and calibration fibre (B), which is at 7$^{\prime\prime}$ from A, on sky.

The spectra were reduced using the dedicated data reduction pipeline (version {\tt 2.3.3}), provided by the ESPRESSO consortium and ESO, and run on the {\tt esoreflex} environment. Briefly, the reduction cascade includes bias and dark subtraction, flat-field correction, slice identification and wavelength calibration. For the purpose of solving the dispersion solution, day-time calibration frames taken with the Thorium-Argon lamp are used. We chose not to use the sky-subtracted spectra as lunar contamination in the science spectra is expected to be negligible due to its phase and angular distance (41\% at 114\,deg) and given the magnitude of the target ($m_V = 9.82$), thereby avoiding an additional source of noise in the final reduced spectra.

\begin{figure}[t]
    \centering
    \includegraphics[width=\linewidth]{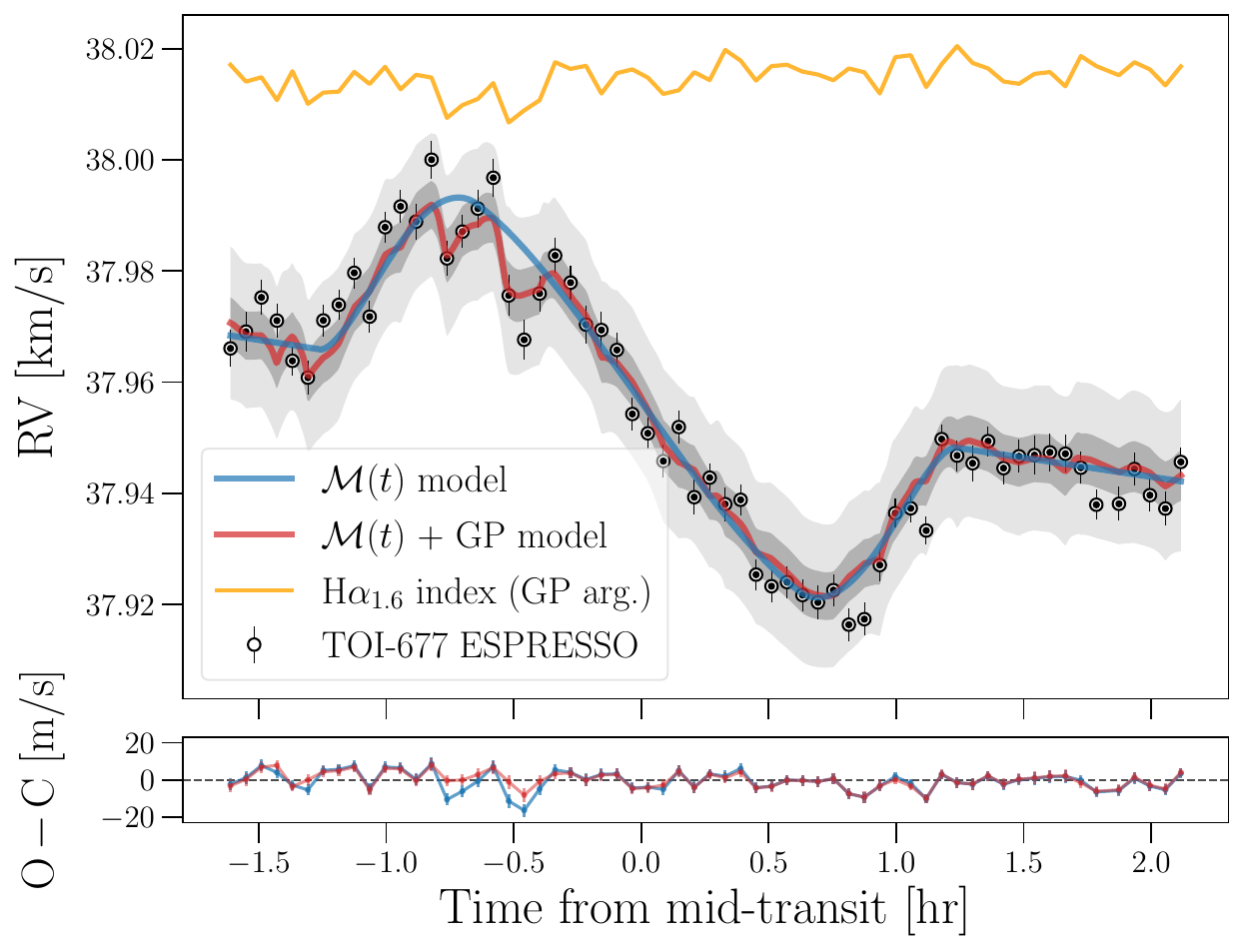}
    \caption{\textit{Top:} Radial velocity measurements of TOI-677 during the primary transit of TOI-677\,b, demonstrating the RM effect, presented as circles including errorbars. Note that neither the systemic velocity nor the underlying orbital reflex motion have been removed and have been included as parameters in the modeling process. The blue line represents the best fit analytical model $\mathcal{M}(t)$, including an RM and RV component, and the red line is this same model which also includes the GP noise model component. The dark and light gray shaded regions represent the 1 and 3$\sigma$ uncertainties of this mode, respectively. The orange line shows the variations of the H$\alpha_{1.6}$ index, where the units have not been included (see bottom panel of Figure \ref{fig:observations}). This represents the free regressor used to model the covariance matrix. \textit{Bottom:} the residuals of both models, where colour have been chosen according to the relevant model.}
    \label{fig:final fit}
\end{figure}

The pipeline also calculates the cross-correlation function (CCF) of the spectra with a binary mask for the stellar type matching closest the spectral type of the observed target (F9 in our case). We calculated the CCF at steps of 0.5\,km/s, for $\pm$\,40\,km/s centred on the expected systemic velocity of the star. The CCF from individual slices are summed (excluding those slices heavily contaminated by telluric absorption lines) and a Gaussian fit to this final CCF determines the central position of the profile and therefore the radial velocity. These calculated radial velocities together with their respective uncertainties, are presented in Table \ref{tab:RVs} (Appendix \ref{sec:Ap1}) and demonstrated in Figure \ref{fig:final fit}, where the RM anomaly is clearly evident. Additionally, the pipeline provides S/N calculations for the individual spectral orders (middle panel of Figure \ref{fig:observations}), as well as a series of diagnostics determined from the CCF, which we used to search for correlations with the residuals of our eventual model. Further to the data reduction pipeline, we also used the dedicated Data Analysis Software (DAS, version {\tt 1.3.3}) to determine activity indices from the spectra, such as the S-index and $\log R^{\prime}_{\textrm{HK}}$, the latter of which is shown in the bottom panel of Figure \ref{fig:observations}.

\section{Data analysis} \label{sec:analysis}

It has been shown that depending on the methodology through which the radial velocities are extracted from the observed spectra, one obtains different shapes for the RM effect \citep{Boue2013}. As we obtained RV values for TOI-677 through the fitting of a Gaussian function to the CCF, we use the publicly available code {\tt ARoME} \citep{Boue2013} to model the RM effect, as this code provides instantaneous RM function definitions for RVs estimated through the cross-correlation and iodine cell techniques, as well as the weighted mean method. In the function definition, for the treatment of the stellar limb darkening, we use the quadratic law \citep{Kopal1950}. We calculate the Cartesian coordinates of the planet at a given observation time as:
\begin{eqnarray}
x(t) &=& r(t)(\cos{\lambda} \cos{u(t)} - \sin{\lambda} \sin{u(t)} \cos{i})\\
y(t) &=& r(t)(\sin{\lambda} \cos{u(t)} + \cos{\lambda} \sin{u(t)} \cos{i})\\
z(t) &=& r(t)(\sin{u(t)} \sin{i}) \\
\textrm{with } r(t) &=& \frac{a(1-e^2)}{1+e\cos{\nu(t)}} \\
\textrm{and } u(t) &=& \omega + \nu(t)
\end{eqnarray}
\noindent where $\lambda$ is the sky-projected obliquity angle, $i$ is the orbital inclination angle, $a$ is the orbital semi-major axis scaled to the stellar radius, $e$ is the orbital eccentricity, $\nu(t)$ is the true anomaly, $\omega$ is the argument of the periapsis, \textit{u} is the argument of latitude (not to be confused with the limb darkening coefficients) and \textit{r} is the radius from true anomaly. The \textit{x} and \textit{y} axes point along the plane of the sky (pointing arbitrarily to the right and up, respectively) and the \textit{z} axis towards the observer.

Once the position of the planet is defined at each time of observation, the anomalous radial velocity value is calculated using the RM model, introduced above. The final model $\mathcal{M}$ is subsequently the sum of the underlying RV trend of the star, which has a systemic and a planetary component, and this RM anomaly:
\begin{equation}
\begin{aligned}
    \mathcal{M}(t) = {} & \gamma_0 + \textrm{RV}_{\textrm{orb}}(t,P,K,e,T_0,\omega) \\
                  & ~~~+ \textrm{RV}_{\textrm{RM}}(t,T_0,a,P,e,i,\nu\sin{i_\star},\\
                  & {~~~~~~~~~~~~~~~~} \omega,\lambda,u_1,u_2,R_p,\sigma_0,\beta_0,\zeta)    
\end{aligned}
\end{equation}
\noindent where $\gamma_0$ is the systemic velocity, $P$ is the orbital period, $K$ is the RV semi-amplitude, $T_0$ is the time of mid-transit, $u_1$ and $u_2$ are the limb darkening coefficients, $R_p$ is the planetary radius scaled to the stellar radius, $\sigma_0$ is the width of the CCF (FWHM of a Gaussian fit to the CCF), $\beta_0$ is the line-width of the non-rotating star and $\zeta$ is the stellar macro-turbulence velocity.

\subsection{Noise consideration}
\begin{deluxetable*}{lccc}[t]
\tabletypesize{\small}
\tablewidth{0pt} 
%\tablenum{1}
\tablecaption{Best fit parameter values from the MCMC simulations. \label{tab:results}}
\tablehead{
\colhead{Parameter} & \colhead{Prior\tablenotemark{a}} & \colhead{\citet{Jordan2020}} & \colhead{This work} }
%\colnumbers
\startdata 
Mid-transit time, T$_0$\,[$-2459558$\,BJD$_{\textrm{TDB}}$]                 & $\mathcal{U}(0.7191,0.8191)$ & -- & $0.769306^{+0.000700}_{-0.000655}$ \\
Orbital Period, P\,[days]                         & $\mathcal{N}(11.2366,0.00011^2)$ & 11.23660\,$\pm$\,0.00011 & $11.23660 \pm 0.00011$ \\
Orbital eccentricity, $e$            & $\mathcal{N}(0.435,0.024^2)$ & 0.435\,$\pm$\,0.024 & $0.443 \pm 0.021$ \\
Argument of periastron, $\omega$\,[rad]           & $\mathcal{N}(1.23,0.06^2)$ & 1.23\,$\pm$\,0.06 & $1.23 \pm 0.06$ \\
RV semi-amplitude, $K$\,[m/s]                     & $\mathcal{N}(111.6,0.5^2)$ & -- & $111.6 \pm 0.5$ \\
Systemic velocity, $\gamma_0$\,[$-37$\,km/s]               & $\mathcal{U}(0.935,0.945)$ & -- & 0.94068$^{+0.00497}_ {-0.00535}$\\
Scaled semi-major axis, $a/R_\star$                        & $\mathcal{U}(12,25)$ & 17.44\,$\pm$\,0.69 & 15.86$^{+1.58}_{-1.32}$ \\
Relative planetary radius, $R_p/R_\star$                   & $\mathcal{N}(0.0942,0.0012^2)$ & 0.0942$^{+0.0010}_{-0.0012}$ & $0.0942\pm0.0012$ \\
Orbital inclination, $i$\,[deg]                            & $\mathcal{U}(80,90)$ & 85.86$^{+0.11}_{-0.10}$\tablenotemark{b} & 84.80$^{+0.80}_{-0.79}$ \\
Orbital impact parameter, $b$ (derived) & -- & $0.723^{+0.018}_{-0.024}$ & $0.858^{+0.272}_{-0.220}$\\
Sky-projected obliquity, $\lambda$\,[deg]         & $\mathcal{U}(-180,180)$ & -- & $0.3\pm1.3$ \\
Equatorial stellar rotation, $\nu\sin{i_\star}$\,[km/s]    & $\mathcal{U}(0,15)$ & 7.80\,$\pm$\,0.19 & 6.91$^{+1.32}_{-1.20}$ \\
Stellar macro-turbulence velocity, $\zeta$\,[km/s]         & $\mathcal{N}(5.5,0.5^2)$ & -- & 5.53$^{+0.50}_{-0.51}$ \\
Linear limb darkening coefficient, $u_1$                   & -- & 0.50$_{\textrm{TESS}}$ & 0.5153 (fixed) \\
Quadratic limb darkening coefficient, $u_2$                & -- & $-0.06_{\textrm{TESS}}$ & 0.1518 (fixed) \\
GP kernel amplitude, $\mathcal{A}$\,[m/s]                  & $\Gamma(1,0.1)$ & -- & $6.4^{+2.3}_{-1.7}$ \\
GP kernel regressor length scale, $\ell$                   & $\Gamma(1,0.1)$ & -- & 0.0067$^{+0.0022}_{-0.0028}$ \\
White noise, $\sigma_w$\,[m/s]                             & $\mathcal{U}(0,\infty)$ & -- & 4.2$^{+0.5}_{-0.4}$
\enddata
\tablenotetext{a}{The distributions in the prior column are defined as: $\mathcal{U}(l,u)$ is a uniform distribution between $l$ and $u$, $\mathcal{N}(\mu,\sigma^2)$ is a normal distribution with a mean of $\mu$ and a variance of $\sigma^2$, and $\Gamma(k,\theta)$ is a gamma distribution with a shape parameter $k$ and a scale parameter $\theta$.}\tablenotetext{b}{Inclination was reported incorrectly in \citet{Jordan2020} as $i=87.63$\,deg. The fitted parameter in that work was actually $b$, whose reported value is correct. The reported inclination was derived incorrectly using the relation between $i$ and $b$ valid only for a circular orbit.}
%\tablecomments{placeholder.}
\end{deluxetable*}

\begin{figure}[t]
    \centering
    \includegraphics[width=\linewidth]{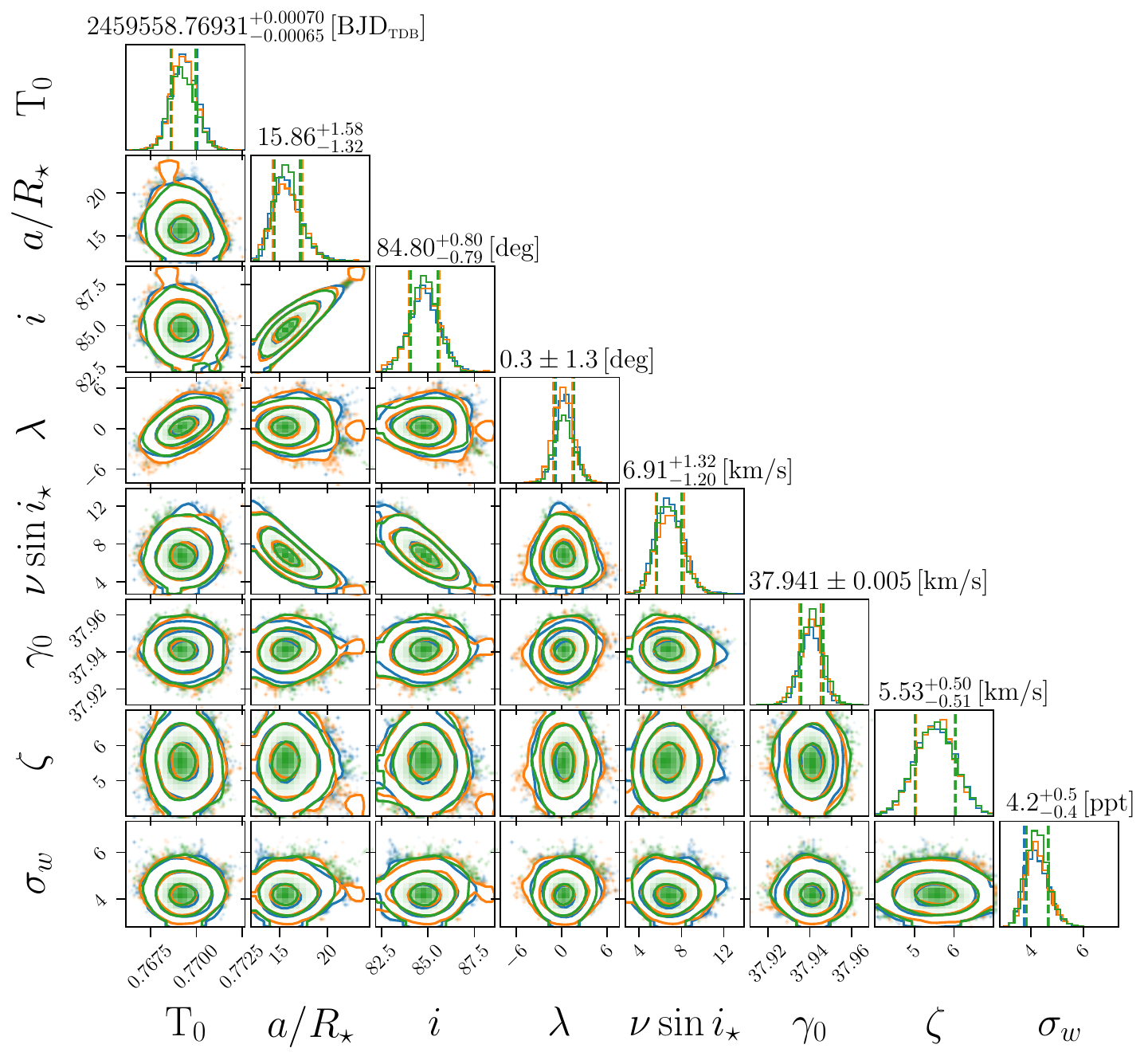}
    \caption{Joint and singular posterior probability distributions for some of the fitted parameters of the GP model $\mathcal{M}(t)$ fitted to the ESPRESSO transit RV data of TOI-677\,b. The different colours represent the 3 independent MCMC simulations, where the lines show the 1, 2 and 3$\sigma$ levels of the joint distributions. At the top of each column, the median and the 16th/84th quartiles of the combined distributions are given as the final best-fit solution for each parameter, with the full results for all the parameters given in Table \ref{tab:results}.}
    \label{fig:posteriors}
\end{figure}
We initially fitted the data with this analytical model, assuming only uncorrelated noise.  However, the residuals of this fit, not presented in this manuscript, presented a distribution clearly deviating from the expected Gaussian. This points to the presence of correlated noise in the observations, caused by astrophysical and/or instrumental sources. The presence of active regions on the stellar surface has been shown to introduce anomalies in the observed photometric light curves both in and out of transit \citep{Rackham2018}, as well as in the radial velocity measurements \citep{Huerta2008}. We therefore measured a series of activity indices from our observed spectra, including the FWHM, bisector span and contrast of the CCF, $\log{R^{\prime}_{\textrm{HK}}}$ and the S-index, as well as line indices for H$\alpha_{0.6}$, H$\alpha_{1.6}$\footnote{The subscripts indicate the widths of the central bandpasses used to calculate the index.}, He\,{\small I} and Na\,{\small I} lines \citep{Gomes2011}. Two of these indices are presented in the bottom panel of Figure \ref{fig:observations}. Furthermore, the Ca\,{\small I} activity-insensitive line index was also calculated as control. The estimation of these line indices was made using the {\tt ACTIN} python package \citep{Gomes2018}. Of all these indices, the variations in the H$\alpha_{1.6}$ index present the only clear sign of correlation with the residuals. This was searched for visually, as well as with a simple correlation analysis, this index showing a significant correlation (Pearson's correlation coefficient of 0.86). Namely, the two sharp decreases in this index at approximately 0.75 and 0.5 hours before mid-transit (indicated with red arrows in the bottom panel of Figure \ref{fig:observations}), coincide with RV deviations from the noise-free model. We checked for possible contamination of the H$\alpha$ line with mirco-telluric absorption lines, whose variation could mimic line index variability. To this end we modelled the telluric absorption in the spectral series with ESO's {\tt molecfit} \citep[v. 4.2.3;][]{Smette2015,Kausch2015}, and note that those telluric lines, included in the region used for the calculation of the H$\alpha_{1.6}$ index, are not responsible for the variability observed. The calculated indices together with their uncertainties are presented in Table \ref{tab:RVs}.

We incorporate these index measurements into our model via a Gaussian Process (GP). The covariance matrix $\Sigma$ of the GP is modelled with a squared exponential kernel:
\begin{equation}
    \Sigma_{ij} = \mathcal{A} \exp{ \left( \frac{-(\textrm{H}\alpha_i-\textrm{H}\alpha_j)^2}{2\ell^2} \right)} + \delta_{ij}\sigma^2,
\end{equation}
with H$\alpha$ being the line indices measured for the 1.6\,\AA\,bandpass, $\mathcal{A}$ and $\ell$ the kernel amplitude and the length scale, respectively, and $\sigma$ the uncorrelated or white noise in the data. The implementation of this GP noise model is performed with the {\tt GeePea} python module \citep{Gibson2012}. 

To sample the posterior distributions we ran 3 independent MCMC simulations of 120\,000 steps each, using an Affine invariant ensemble sampler \citep{Goodman2010}, assuming restrictive Gaussian prior distribution for the stellar macro-turbulence velocity\footnote{We initially tried to fit for this parameter without a restrictive prior, but convergence was not achieved. Subsequently, the prior distribution is drawn from the relation estimated by \citet{Doyle2014} using astroseismic rotational velocities from \textit{Kepler} data.} ($\zeta$), as well as the orbital period $(P)$, the eccentricity $(e)$, the argument of periastron $(\omega)$, the RV semi-amplitude $(K)$ and the relative planetary radius $R_p/R_\star$, whose values were taken from \citet{Jordan2020} through the analysis of TESS light curves and RV monitoring data. This approach was taken to ensure that the uncertainties on those parameters are correctly propagated. However, one caveat to note is that such restrictive Gaussian priors do not account for the impact of the existing correlations between the scaled semi-major axis, $\nu\sin{i_\star}$ and the eccentricity, and therefore the quoted uncertainties could be slightly underestimated. The priors are detailed in Table \ref{tab:results}. The two coefficients of the quadratic limb-darkening law are fixed to those calculated from PHOENIX stellar spectrum model library of \citet{Husser2013}, for the ESPRESSO bandpass using {\tt PyLDTK} \citep{Parviainen2015}. For all other model parameters we assumed flat, uninformative prior distributions details of which are presented in Table \ref{tab:results}. Additionally, for the kernel parameters we assumed very restrictive gamma priors with the shape parameter equal to 1 to maximize the distribution at 0 and the scale parameter to 0.1 in order to encourage the probability distributions to converge towards 0. This approach ensures that the included GP regressor contributes to the covariance only when there exists a significant correlation with the systematic noise present.

The best fit analytical model, as well as the noise model, together with their residuals are presented in Figure \ref{fig:final fit}. The posterior probability distributions and the joint posteriors are presented in Figure \ref{fig:posteriors}, with the independent chains over-plotted (the initial 20\,000 steps of which are burnt in). The best fit results for the fitted parameters, given in Table \ref{tab:results}, are derived from the median and the 16th and 84th percentiles of those distributions. We obtain a perfectly aligned sky-projected orbit of TOI-677\,b with respect to the spin orbit of its host star, with $\lambda = 0.3\pm1.3$\,deg. All other estimated parameters are in general agreement with previously obtained results, whereby our parameters result in a slightly more inclined orbit.

\subsubsection{True obliquity}
As the angle measured from this analysis of the RM effect is the sky-projected ($\lambda$) portion of the true obliquity angle ($\psi$), we attempted to estimate this true value. One can potentially de-project this measurement if the stellar line of sight inclination ($i_\star$) can be measured via $\sin^{-1} \left( \nu\sin i_\star / (2\pi R_\star/P_{\textrm{rot}}) \right)$, although this approach suffers from biases due to the fact that $\nu \sin{i_\star}$ and 2$\pi R_\star/P_{\textrm{rot}}$ ($\equiv \nu$) are not statistically independent measurements \citep{Morton2014,Masuda2020}. $\psi$ would subsequently be estimated through its geometrical relation to the planetary orbital plane inclination ($i$) and $i_\star$:
\begin{equation}
    \cos{\psi} = \cos{i_\star}\cos{i} + \sin{i_\star}\sin{i}\cos{\lambda}
\end{equation}

To this effect, we attempted to measure the stellar rotation period ($P_{\textrm{rot}}$) through modulations in the TESS light curves, both the simple aperture photometry (SAP) and the pre-search data conditioning (PDC) LCs, imprinted by the rotation of active regions on the star.  However, a Lomb-Scargle periodogram search of all available observations of TOI-677 from TESS sectors 9, 10, 35 and 36 did not result in a viable detection.

\begin{deluxetable*}{lccc}[t]
\tabletypesize{\small}
\tablewidth{0pt} 
%\tablenum{1}
\tablecaption{Best fit parameter values from the nested sampling simulations for the two-planet fit of the RV data, assuming both circular and eccentric orbits for the possible outer companion. \label{tab:two planet fit results}}
\tablehead{
\colhead{Parameter} & \colhead{Prior} & \colhead{Circular Orbit} & \colhead{Eccentric Orbit}}
%\colnumbers
\startdata 
Mid-transit, T$_{0,1}$\,$-2458547$\,[BJD] & $\mathcal{N}(0.4743,0.0012)$ & $0.474303^{+0.001118}_{-0.001133}$ & $0.474312^{+0.001060}_{-0.001045}$ \\
Period, P$_1$\,[days] & $\mathcal{U}(11.1,11.5)$ & $11.236518^{+0.000589}_{-0.000609}$ & $11.236168 \pm 0.000537$ \\
Eccentricity, $e_1$ & $\mathcal{N}(0.434,0.05)$ & $0.442 \pm 0.018$ & $0.436 \pm 0.019$ \\
Argument of periastron, $\omega_1$\,[deg] & $\mathcal{N}(70.47,1.0)$ & $70.47 \pm 0.09$ & $70.47 \pm 0.44$\\
RV semi-amplitude (planet 1), $K_{\star,1}$\,[m/s] & $\mathcal{N}(111.6,5.0)$ & $111.59 \pm 2.84$ & $113.48^{+2.74}_{-2.64}$\\ 
Mass, $m_{p,1}$\,[$M_{\textrm{jup}}$] & $-$ & $1.24 \pm 0.13$ & $1.26 \pm 0.14$\\ \hline
Mid-transit, T$_{0,2}$\,$-2460000$\,[BJD] & $\mathcal{U}(0,10^4)$ & $579.329132^{+53.885799}_{-61.530429}$ & $1755.376054^{+392.459980}_{-549.826780}$ \\
Period, P$_2$\,[days] & $\mathcal{U}(11.5,10^4)$ & $\gtrsim 3953.751212^{+111.367231}_{-127.364290}$ & $\gtrsim 4901.138760^{+523.775478}_{-644.759743}$ \\
Eccentricity, $e_2$ & $\mathcal{U}(0,1)$ & 0 (fixed) & $0.436^{+0.067}_{-0.058}$ \\
Argument of periastron, $\omega_2$\,[deg] & $\mathcal{U}(-180,180)$ & 90 (fixed) & $-123.68^{+6.96}_{-6.99}$ \\
RV semi-amplitude (planet 2), $K_{\star,2}$\,[m/s] & $\mathcal{U}(0,1000)$ & $450.98^{+12.39}_{-13.46}$ & $594.91^{+35.01}_{-35.34}$\\
Mass lower limit, $m_{p,2}\sin{i}$\,[$M_{\textrm{jup}}$] & $-$ & $\gtrsim 39.20 \pm 2.81$ & $\gtrsim 49.99 \pm 14.07$\\ \hline
Log evidence, $\ln{\mathcal{Z}}$ & $-$ & $-683.72 \pm 0.18$ & $-675.04 \pm 0.04$\\
\enddata
\tablecomments{Subscript 1 refers to the inner planet and 2 to the outer companion. The period, and consequently the lower mass limit, of the possible outer companion are presented only as a lower limits since the fitted orbit is not closed.}
\end{deluxetable*}

\begin{figure*}[t]
    \centering
    \includegraphics[width=\linewidth]{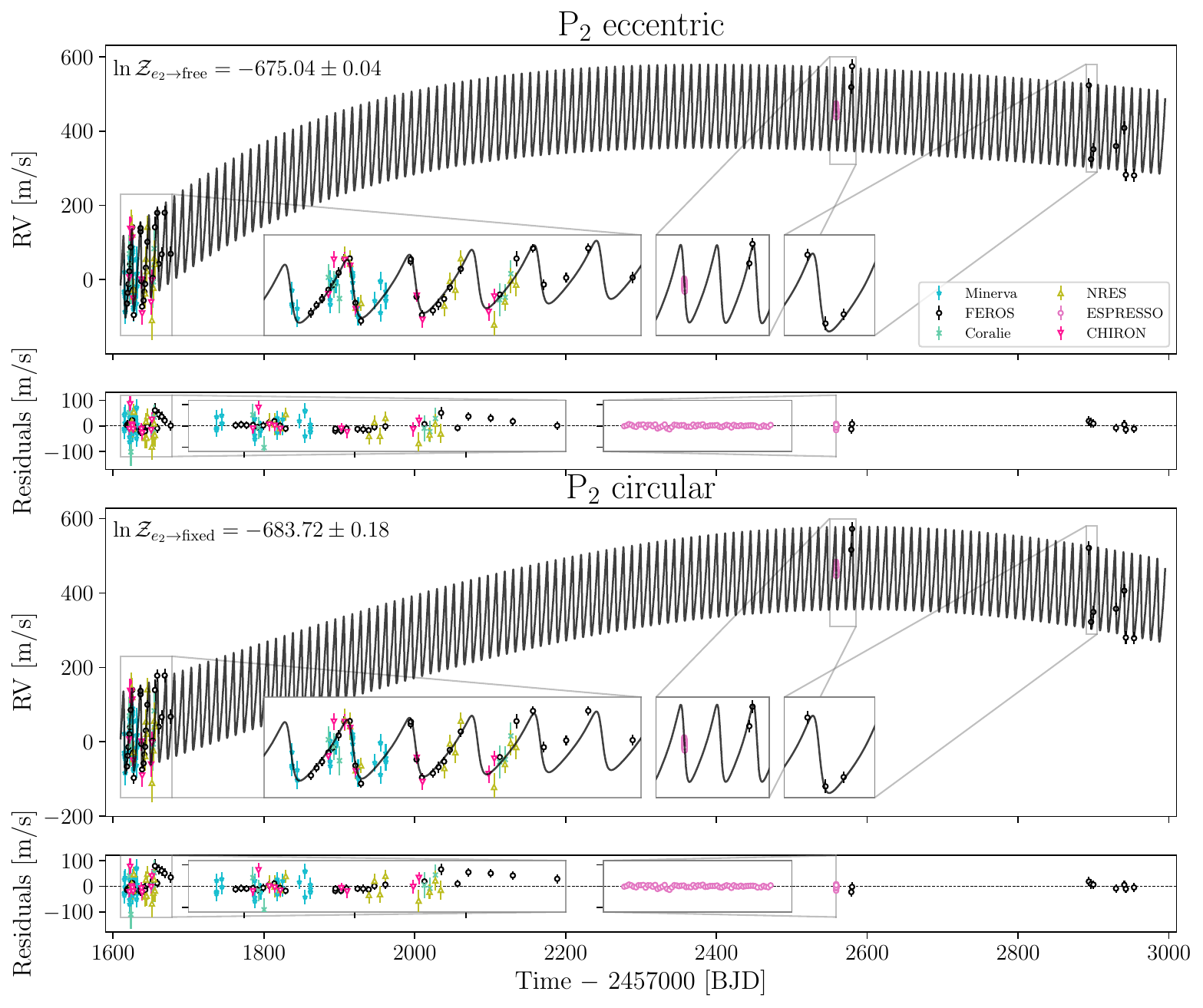}
    \caption{\textit{Top:} TOI-677 RV data from all the different instruments used for measurement. The solid black line represents the best fit model where the eccentricity is taken as a free parameter.  The posteriors of this fit are given in blue in bottom part of Figure \ref{fig:two planet fit posteriors}.  The smaller panel below shows the residuals of this fit.  \textit{Bottom:} is the same as the top, but with a model fit with the eccentricity fixed to 0.  In all panels, zooms have been made into the regions of dense data for better demonstration of the model precision.  For both cases, the log evidence is annotated at the top left corner.}
    \label{fig:two planet fit}
\end{figure*}

\subsection{Possible outer companion}
In the analysis of TOI-677 RV measurements, \citet{Jordan2020} detected an underlying slope of 1.58\,$\pm$\,0.19\,m\,s$^{-1}$day$^{-1}$.  In order to investigate possible roots of such trend in the data, in addition to the ESPRESSO data presented previously, we also observed TOI-677 with FEROS (the Fiber-fed Extended Range Optical Spectrograph), mounted at the MPG/ESO 2.2m telescope at La Silla observatory, in nine distinct epochs.  The stellar RVs were subsequently derived from the spectra via processing with the {\tt CERES} pipeline \citep{Brahm2017b}, similar to what was performed in \citet{Jordan2020}.  These additional radial velocities are given in Table \ref{tab:new RVs} in Appendix \ref{sec:Ap1}, which together with the initial RV data of \citet{Jordan2020} and the ESPRESSO data presented in this work\footnote{ESPRESSO RVs included for this analysis had first the RM effect subtracted, leaving only variations due to the stellar reflex motion present, which are plotted as pink data points in Figure \ref{fig:two planet fit}.}, are used to search for possible outer companions to TOI-677\,b.

\begin{figure}[t]
    \centering
    \includegraphics[width=\linewidth]{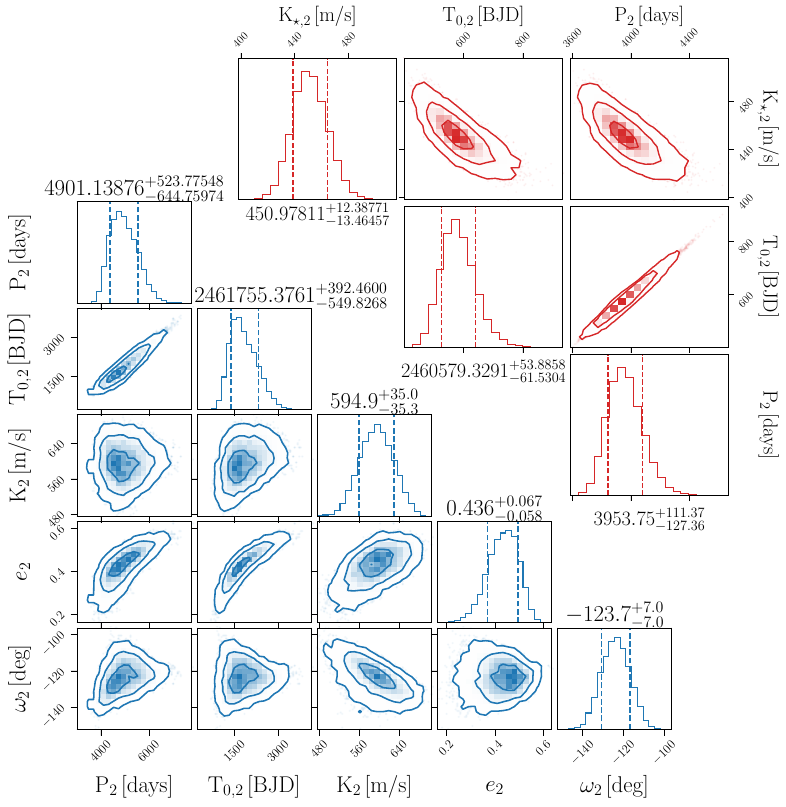}
    \caption{\textit{Top:} Posterior probability distributions from the Nested Sampling algorithm, for the fitted orbital parameters of the possible outer companion to TOI-677\,b.  The blue plot (bottom left) is for the fit with eccentricity taken as a free parameter, and the red plot (top right) for a circular orbit fit.  For each posterior probability distribution, the median and the upper and lower 68th percentile confidence intervals are given as solutions and plotted as dashed lines.}
    \label{fig:two planet fit posteriors}
\end{figure}

We analyze this newly assembled RV data using the {\tt juliet} package \citep{Espinoza2019}, which utilises Keplerian orbital radial velocity perturbation formalism via the {\tt radvel} package \citep{Fulton2018}. In contrast to the model fit performed by \citet{Jordan2020}, instead of an underlying linear trend, we include a second body inducing the long period trend observed in the data (Figure \ref{fig:two planet fit}). We performed two separate fits to the data, whereby the outer component is assumed to be on either a circular or eccentric orbit. In both scenarios, we fitted for orbital parameters of both bodies, as well as instrumental parameters. We found the instrumental dependent systemic velocity ($\gamma$), as well as instrumental jitter ($\eta$) values consistent with those reported by \citet{Jordan2020}, in both sets of analyses. We sampled the Bayesian posterior distributions using the importance nested sampling and {\tt MultiNest} algorithms \citep{Feroz2019}, implemented by {\tt juliet} via the {\tt PyMultiNest} python package \citep{Buchner2014}, using 2000 live points. The two sets of posterior co-distributions and probability distribution functions are presented in Figure \ref{fig:two planet fit posteriors}, where orbital parameters for only the possible outer companion are presented. It must be noted that we do not observe any other significant peaks in the posterior distributions of any of the parameters in either fit. All derived orbital parameters for both bodies in the system, from both modelling approaches, have been presented in Table \ref{tab:two planet fit results}, with the best fit models and their respective residuals shown in Figure \ref{fig:two planet fit}.

In order to evaluate the statistical significance of the two-body model as compared to the single-planet case, we also fitted all the RV data assuming just one planet in the system, which results in $\Delta \ln{\mathcal{Z}} \sim 1500$, as compared to the two-body scenarios, i.e. pointing to a significant preference of the data for the two-body model and the possible presence of an outer companion. Furthermore, there is also strong preference for an eccentric orbit of this possible outer companion, as compared to a circular orbit, from the ratio of the likelihoods of the two models, with $\Delta \ln{\mathcal{Z}} = 8.68$ in its favour. This points to a \textit{very strong} ($2\Delta \ln{\mathcal{Z}} > 10$) preference for the two-body model with an outer companion on an eccentric ($e = 0.44 \pm 0.07$) and very wide orbit of $\gtrsim\,4901$\,days ($\gtrsim\,13.4$\,yr) period. These estimated period values are taken only as lower limits since the orbit is not closed. The lower limit for the mass of this possible outer companion is estimated as $\sim$\,39 and $\sim$\,50$\,M_{\mathrm{jup}}$ for the circular and eccentric cases, respectively, putting it in the brown dwarf regime in either case. Assuming the outer companion is on a relatively coplanar orbit to the inner companion, the true mass of this outer companion is likely close to this lower limit.

However, it must be stressed that such analysis only points to the possible presence of the outer companion, since the data simply do not cover a long enough baseline for any definitive conclusions to be made.  This fact is reflected in the relatively large uncertainties in the determination of the orbital parameters for this potential outer companion, as presented in Table \ref{tab:two planet fit results}.

Additionally, we attempted to fit the RV data with a 3-planet model, keeping $e$ and $\omega$ as free parameters, however convergence was not achieved as the stopping criterion for the nested sampling algorithm could not be reached. The final $\ln{\mathcal{Z}}$ at the moment of stopping the algorithm still pointed to a very strong preference for the two planet model.

%\clearpage

\begin{figure*}[ht]
    \centering
    \includegraphics[width=\linewidth]{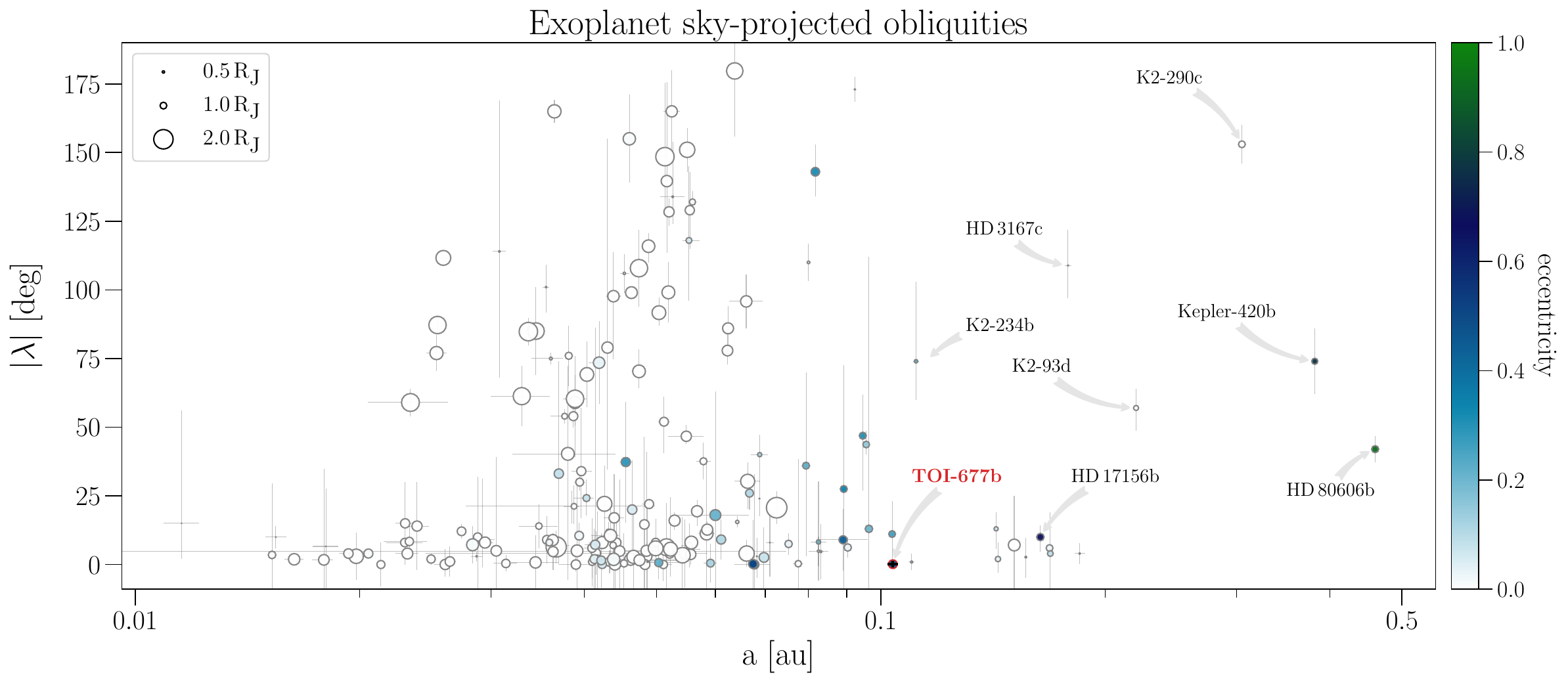}
    \caption{Sample of all currently measured absolute sky-projected obliquity angles $\lambda$ for exoplanets as a function of semi-major axes. The colours of the data points represent the eccentricity of each system, and the sizes are indicative of planetary radius. The data point representing TOI-677\,b is highlighted with a thick red border and text, while all warm planets on misaligned orbits have been annotated in black. Values used for this plot are compiled from the TEPCat \citep{Southworth2011} database.}
    \label{fig:obliquities}
\end{figure*}

\section{Discussion} \label{sec:discussion}
TOI-677\,b is now one of $\sim$\,200 exoplanets for which the projected stellar spin-orbit misalignment $\lambda$ has been measured\footnote{From the TEPCAT catalog \citep{Southworth2011}, which can be found at \url{https://www.astro.keele.ac.uk/jkt/tepcat/}.}. Nearly $85\%$ of these systems correspond to close-in gas giants ($R_{\rm p}>0.7~R_{\rm J}$, $P\leq200$~d), of which $\sim90\%$ are ``hot'' ($P<10$~d) and $\sim10\%$ are ``warm''.  We show the distribution of $|\lambda|$ in Figure \ref{fig:obliquities}, plotted as a function of planet semi-major axis, where the symbol sizes scale with planet radii, and the color scale represents orbital eccentricity. From the Figure, one can distinguish the hot population from the warm population: in the former case, obliquities are distributed broadly \citep[e.g.][]{Fabrycky2009,Morton2014,Munoz2018}, whereas in the latter case, obliquities are distributed rather narrowly \citep{Rice2022}.  In addition, as it is well known, the most compact orbits have zero eccentricity, an indication of circularization owing to tidal dissipation in the planet \citep[e.g.,][]{Goldreich1966}. Since the circularization rate is a steep function of separation \citep{Goldreich1966, Hut1981}, wider orbits may allow for non-zero eccentricity. Indeed, several warm Jupiters have eccentricities above 0.4:
e.g., HD\,80606\,b \citep[$e\approx0.93$;][]{Naef2001},
Corot-10\,b \citep[$e\approx0.53$;][]{Bonomo2010},
Kepler-419\,b \citep[$e\approx0.85$;][]{Dawson2014},
Kepler-420\,b \citep[$e\approx0.77$;][]{Santerne2014},
or TOI-2179\,b \citep[$e\approx0.58$;][]{Schlecker2020}. 

We do not expect TOI-677\,b to have been fully circularized
over the lifetime of its host star ($\approx3$\,Gyr; \citealp{Jordan2020}). Indeed, assuming that tidal dissipation takes place primarily within the planet,
the characteristic circularization timescale is given by:
\begin{equation}\label{eq:gamma}
\tau_{\rm circ}\equiv -\frac{e}{\dot{e}}=
\frac{2F(e)}{7}\tau_{\rm dec} 
\;\;\;\text{with}\;\;\;
\tau_{\rm dec}\equiv \frac{P}{9\pi}{Q_{\rm p}'}\frac{m_{\rm p}}{M_\star}\left(\frac{a}{R_{\rm p}}\right)^5
\end{equation}
\citep{Goldreich1966}, where $\tau_{\rm dec}$ is the characteristic orbital decay timescale, $Q_{\rm p}'$ is the planet's modified tidal quality factor \citep[e.g.,][]{Goldreich1966,Ogilvie2007} and $F(e)$ is an eccentricity-dependent correction factor \citep{Hut1981}. For $e=0.435$, and assuming that the planet is in pseudo-synchronous rotation\footnote{The timescale for the planet's tidal realignment, under weak friction theory, is given by $\tau_{\rm align}\simeq 2(S_{\rm p}/L_{\rm p})\tau_{\rm dec}$ \citep{Hut1981}, where $S_{\rm p}$ is the planet's spin angular momentum, and
$L_{\rm p}=m_{\rm p}\sqrt{GM_\star a (1-e^2)} \approx 3\times10^{42}$\,m$^2$\,kg\,s$^{-1}$ is the planet's orbital angular momentum. If we assume that $S_{\rm p}\sim S_{\rm J}$, with $S_{\rm J}\sim 10^{38}$\,m$^2$\,kg\,s$^{-1}$ being the spin of Jupiter \citep[e.g.,][]{Helled2011}, then the assumption of pseudo-synchronization is well justified.},
we have $F(e)\approx0.2$.
 Further assuming that $Q_{\rm p}'=10^5-10^6$ \citep[e.g.,][]{Goldreich1966,Yoder1981}, we have $\tau_{\rm circ}\sim 1-10$\,Gyr for TOI-677\,b. Had we not ignored tidal dissipation in the star\footnote{%%%%%%%%%%%%%%%%%%%%%%%%%%%%%%%%%%%%%%%%%%%%%%%%%%%%%%%%%%%%%%%%%%%%%%%%%%%%%%%%%%
Tidal dissipation due to planetary tides {\it on the star} contributes to the circularization rate by a factor $Q_{\rm p}'/Q_\star'(m_{\rm p}/M_\star)^2(R_\star/R_{\rm p})^5$ times smaller, and thus, unless $Q_\star'\ll Q_{\rm p}'$, it can be safely neglected \citep[e.g.,][]{Matsumura2008}.},
%%%%%%%%%%%%%%%%%%%%%%%%%%%%%%%%%%%%%%%%%%%%%%%%%%%%%%%%%%%%%%%%%%%%%%%%%%%%%%%%%%%%%%%%%%%%%%%%
these timescales would be negligibly shorter for any value of $Q_\star'$ greater than $10^7$, which is to be expected in this type of system \citep[e.g.,][]{Barker2010,Barker2011,Penev2011}. Similarly, for such values of $Q_\star'$, tidal realignment of the star itself would  take hundreds of times longer than the age of the system. Moreover, even if stellar realignment did take place, it would come at the expense of planetary engulfment \citep{Barker2009}. Thus, only a tidal theory that goes well beyond weak friction \citep[e.g.,][]{Lai2012} could possibly permit the realignment of the stellar spin while sparing the planet's orbit.

TOI-677\,b belongs to an intriguing, emerging group of eccentric, spin-orbit-aligned systems bracketed between the hot and warm populations. At these orbital separations, dissipation of energy within the star--responsible for obliquity damping \citep{Hut1981}--is extremely weak, which rebuffs the hypothesis of tidal reprocessing of the spin-orbit angle over long timescales \citep[e.g., see][]{Winn2010,Albrecht2012}. Instead, planets such as TOI-677\,b are likely to have attained their unusual orbital configurations soon after the planet formation and migration process was finalized.

Given these properties, systems like TOI-677\,b  present a significant challenge to standard theories of planet migration, both for the dynamically hot and dynamically cold scenarios. On the one hand, planets like TOI-677\,b are unlikely to have attained their eccentricities during dynamically cold, disk-driven planet migration. On the other hand, low spin-orbit alignment would also disfavour dynamically hot, eccentricity excitation mechanisms, which are usually accompanied by large changes in inclination. Moreover, the unlikeliness of large-amplitude eccentricity oscillations being responsible for these elongated orbits would in turn reject the notion that TOI-677\,b-like systems are ``failed'' or ``proto'' hot Jupiters \citep[e.g.,][]{Dong2014,Petrovich2016}. In fact, while there are indeed two planetary systems--HD\,80606 and Kepler-420\,-- that appear to be quintessential examples of ongoing (or failed) high-eccentricity migration driven by Lidov-Kozai oscillations \citep{Wu2003}, these planets do not appear to be representative of the warm giant population. They have, in addition to high eccentricity, known binary companions and high obliquities (see Figure \ref{fig:obliquities}).
 
Thus, we must consider the option of {\it in situ} excitation of high eccentricity while at low inclination. This is indeed possible due to an exterior, nearly coplanar perturber of mass $\gtrsim m_{\rm p}$ and of moderate-to-high eccentricity \citep{Lee2003}. If sufficiently eccentric, the candidate sub-stellar mass companion discovered in this study presents an ideal potential explanation for TOI-677\,b's peculiar orbit. In order to excite the planet's eccentricity from a circular orbit to its current value of $e_1=0.435$, the outer perturber must satisfy the approximate condition:
\begin{equation}\label{eq:eccentricity_criteron}
    \frac{e_2}{(1-e_2^2)}\geq\frac{4}{15}\frac{e_1}{1+\tfrac{3}{4}e_1^2}
  \left(\frac{P_2}{P_1}\right)^{2/3}
  \approx 5~.
\end{equation}
proposed by \citet{Petrovich2015}. Being a necessary yet not a sufficient condition \citep[it ignores suppressing effects such as general relativistic precession; e.g.,][]{Liu2015}, equation \ref{eq:eccentricity_criteron} can only provide a lower limit on the required value of $e_2$. The currently estimated eccentricity and period for the outer companion, given in Table \ref{tab:two planet fit results}, fail to satisfy this condition by one order of magnitude.

Consequently, we may conclude that no known mechanism of planet migration can explain the current orbital eccentricity and alignment of TOI-677\,b with the currently known objects in the system. This puzzle highlights the importance of obtaining RM observations of a wider class of exoplanets, pushing the boundaries of high precision spectroscopy.

\section{Summary \& Conclusions} \label{sec:conclusions}
In this study we presented single transit observations of the warm Jupiter TOI-677\,b with the ESPRESSO spectrograph, obtaining the Rossiter McLaughlin effect in order to measure the sky projected obliquity angle $\lambda$. This angle was determined to be $0.3\pm1.3$\,deg, putting the planet on a perfectly aligned orbit with the stellar spin axis. In modelling the effect together with the correlated noise we uncovered a strong correlation with the H$\alpha$ activity index, which was used as the regressor in the calculation of the covariance matrix in the Gaussian Process model. In the analysis, MCMC methods were used to determine parameter uncertainties, while evaluating posterior co-distributions. An attempt was made to measure the true obliquity angle of the system, which was unsuccessful due to the inability to measure the stellar rotation period from TESS photometry, owing to the absence of activity-induced light curve modulations. Follow-up radial velocity monitoring revealed a long-term periodic signal, which together with the initial data from \citet{Jordan2020} was modelled with a two-component Keplerian model. The analysis revealed a significant preference for a companion on an eccentric orbit, as opposed to a circular one. This solution pointed to the possible presence of a companion with a lower mass limit in the brown dwarf regime ($M_\textrm{p}$\,$\approx$\,50\,$M_\textrm{J}$), on a wide ($P$\,$\approx$\,13.4\,yr) and moderately eccentric ($e$\,$\approx$\,0.44) orbit. Posteriors obtained from a nested sampling approach revealed relatively well-constrained distributions, although no definitive conclusion was made about the presence of this outer companion, due to the lack of sufficient coverage of this long orbital period.
We finally discussed the orbital architecture of this system in the context of currently known planet migration mechanisms, and the challenges it poses to them. Namely, while it is likely the system attained its eccentricity through disk migration, the aligned orbit disfavours eccentricity excitation mechanisms. Furthermore, we argued that it is also highly unlikely that the system is a failed or proto hot Jupiter. We finally discussed the possibility of an in situ excitation of the eccentricity by the sub-stellar outer companion. However, with the current and limited analysis, it was concluded that this outer companion does not possess high enough eccentricity to cause the elevated eccentricity in the inner planetary companion. This result, subsequently, highlights the need and the importance of obtaining RM measurements for planets in the warm giant regime, to better test and refine planet migration theories.

\begin{acknowledgments}
E.S., A.J., R.B., and C.P.\ acknowledge support from ANID -- Millennium  Science  Initiative -- ICN12\_009.  
A.J.\ acknowledges additional support from FONDECYT project 1210718. R.B.\ acknowledges support from FONDECYT project 11200751.
C.P.\ acknowledges support from ANID Millennium Science Initiative-ICN12\_009, CATA-Basal AFB-170002, ANID BASAL project FB210003, FONDECYT Regular grant 1210425, CASSACA grant CCJRF2105, and ANID+REC Convocatoria Nacional subvencion a la instalacion en la Academia convocatoria 2020 PAI77200076.
\end{acknowledgments}


\vspace{5mm}
\facilities{VLT/UT1(ESPRESSO), MPG/ESO: 2.2m(FEROS)}


\software{ 
ACTIN \citep{Gomes2018}, 
ARoME \citep{Boue2013}, 
astropy \citep{astropy2022}, 
CERES \citep{Brahm2017}, 
corner \citep{corner},
emcee \citep{Foreman2013}, 
GeePea \citep{Gibson2012}, 
juliet \citep{Espinoza2019}, 
matplotlib \citep{Hunter2007}, 
molecfit \citep{Smette2015}, 
MultiNest \citep{Feroz2019}, 
NumPy \citep{Harris2020}, 
PyLDTK \citep{Parviainen2015}, 
PyMultiNest \citep{Buchner2014}, 
radvel \citep{Fulton2018}, 
SciPy \citep{scipy2020}.
          }



\appendix

\section{ESPRESSO and FEROS Radial Velocities}\label{sec:Ap1}
In this appendix we present the radial velocities, as well as the H$\alpha_{1.6}$ index, measured for the TOI-677 spectra obtained with ESPRESSO in Table \ref{tab:RVs}, as well as the additional RVs obtained with FEROS in Table \ref{tab:new RVs}.

\begin{deluxetable}{lcc}
\tabletypesize{\scriptsize}
\tablewidth{0pt} 
%\tablenum{1}
\tablecaption{ESPRESSO radial velocity measurements and H$\alpha$ activity index for the 1.6\,\AA~bandpass. \label{tab:RVs}}
\tablehead{
\colhead{Time} & \colhead{Radial Velocity} & \colhead{H$\alpha_{1.6}$ index} \vspace{-0.2cm} \\
\colhead{[BJD$_{\textrm{TDB}}$]} & \colhead{[km/s]} & \colhead{}
}
%\colnumbers
\startdata 
2459558.7021948  &  37.96608\,$\pm$\,0.00334  &  0.17175\,$\pm$\,0.00022 \\ 
2459558.7047237  &  37.96910\,$\pm$\,0.00352  &  0.17114\,$\pm$\,0.00023 \\ 
2459558.7072542  &  37.97525\,$\pm$\,0.00316  &  0.17130\,$\pm$\,0.00021 \\ 
2459558.7097828  &  37.97108\,$\pm$\,0.00301  &  0.17048\,$\pm$\,0.00021 \\ 
2459558.7123121  &  37.96384\,$\pm$\,0.00262  &  0.17152\,$\pm$\,0.00019 \\ 
2459558.7148410  &  37.96083\,$\pm$\,0.00304  &  0.17034\,$\pm$\,0.00021 \\ 
2459558.7173617  &  37.97112\,$\pm$\,0.00283  &  0.17075\,$\pm$\,0.00020 \\ 
2459558.7198899  &  37.97391\,$\pm$\,0.00273  &  0.17078\,$\pm$\,0.00019 \\ 
2459558.7224178  &  37.97967\,$\pm$\,0.00280  &  0.17149\,$\pm$\,0.00020 \\ 
2459558.7249359  &  37.97180\,$\pm$\,0.00283  &  0.17106\,$\pm$\,0.00020 \\ 
2459558.7274613  &  37.98788\,$\pm$\,0.00286  &  0.17168\,$\pm$\,0.00020 \\ 
2459558.7299908  &  37.99163\,$\pm$\,0.00300  &  0.17087\,$\pm$\,0.00021 \\ 
2459558.7325242  &  37.98888\,$\pm$\,0.00324  &  0.17139\,$\pm$\,0.00022 \\ 
2459558.7350516  &  38.00003\,$\pm$\,0.00336  &  0.17129\,$\pm$\,0.00022 \\ 
2459558.7375717  &  37.98228\,$\pm$\,0.00319  &  0.16984\,$\pm$\,0.00021 \\ 
2459558.7400964  &  37.98709\,$\pm$\,0.00297  &  0.17030\,$\pm$\,0.00021 \\ 
2459558.7426159  &  37.99123\,$\pm$\,0.00339  &  0.17052\,$\pm$\,0.00022 \\ 
2459558.7451400  &  37.99678\,$\pm$\,0.00337  &  0.17109\,$\pm$\,0.00022 \\ 
2459558.7476639  &  37.97561\,$\pm$\,0.00368  &  0.16967\,$\pm$\,0.00023 \\ 
2459558.7501844  &  37.96764\,$\pm$\,0.00359  &  0.17011\,$\pm$\,0.00023 \\ 
2459558.7527129  &  37.97595\,$\pm$\,0.00324  &  0.17047\,$\pm$\,0.00022 \\ 
2459558.7552439  &  37.98280\,$\pm$\,0.00317  &  0.17184\,$\pm$\,0.00022 \\ 
2459558.7577648  &  37.97791\,$\pm$\,0.00299  &  0.17160\,$\pm$\,0.00021 \\ 
2459558.7602871  &  37.97036\,$\pm$\,0.00345  &  0.17171\,$\pm$\,0.00023 \\ 
2459558.7628070  &  37.96939\,$\pm$\,0.00324  &  0.17071\,$\pm$\,0.00022 \\ 
2459558.7653380  &  37.96580\,$\pm$\,0.00314  &  0.17145\,$\pm$\,0.00021 \\ 
2459558.7678639  &  37.95428\,$\pm$\,0.00295  &  0.17158\,$\pm$\,0.00021 \\ 
2459558.7703958  &  37.95082\,$\pm$\,0.00277  &  0.17129\,$\pm$\,0.00020 \\ 
2459558.7729244  &  37.94581\,$\pm$\,0.00303  &  0.17070\,$\pm$\,0.00021 \\ 
2459558.7754543  &  37.95193\,$\pm$\,0.00300  &  0.17083\,$\pm$\,0.00021 \\ 
2459558.7779786  &  37.93933\,$\pm$\,0.00311  &  0.17148\,$\pm$\,0.00021 \\ 
2459558.7805062  &  37.94283\,$\pm$\,0.00257  &  0.17120\,$\pm$\,0.00019 \\ 
2459558.7830299  &  37.93808\,$\pm$\,0.00305  &  0.17228\,$\pm$\,0.00021 \\ 
2459558.7855602  &  37.93885\,$\pm$\,0.00283  &  0.17190\,$\pm$\,0.00020 \\ 
2459558.7880796  &  37.92539\,$\pm$\,0.00276  &  0.17118\,$\pm$\,0.00020 \\ 
2459558.7906045  &  37.92331\,$\pm$\,0.00288  &  0.17170\,$\pm$\,0.00021 \\ 
2459558.7931330  &  37.92405\,$\pm$\,0.00284  &  0.17175\,$\pm$\,0.00020 \\ 
2459558.7956635  &  37.92172\,$\pm$\,0.00286  &  0.17150\,$\pm$\,0.00020 \\ 
2459558.7981924  &  37.92039\,$\pm$\,0.00298  &  0.17140\,$\pm$\,0.00021 \\ 
2459558.8007128  &  37.92260\,$\pm$\,0.00290  &  0.17119\,$\pm$\,0.00021 \\ 
2459558.8032430  &  37.91639\,$\pm$\,0.00296  &  0.17162\,$\pm$\,0.00021 \\ 
2459558.8057738  &  37.91738\,$\pm$\,0.00296  &  0.17148\,$\pm$\,0.00021 \\ 
2459558.8083037  &  37.92711\,$\pm$\,0.00285  &  0.17072\,$\pm$\,0.00020 \\ 
2459558.8108276  &  37.93645\,$\pm$\,0.00262  &  0.17202\,$\pm$\,0.00019 \\ 
2459558.8133571  &  37.93732\,$\pm$\,0.00252  &  0.17209\,$\pm$\,0.00019 \\ 
2459558.8158835  &  37.93332\,$\pm$\,0.00248  &  0.17095\,$\pm$\,0.00019 \\ 
2459558.8184121  &  37.94977\,$\pm$\,0.00253  &  0.17177\,$\pm$\,0.00019 \\ 
2459558.8209383  &  37.94681\,$\pm$\,0.00278  &  0.17242\,$\pm$\,0.00020 \\ 
2459558.8234680  &  37.94543\,$\pm$\,0.00323  &  0.17181\,$\pm$\,0.00022 \\ 
2459558.8259946  &  37.94940\,$\pm$\,0.00270  &  0.17162\,$\pm$\,0.00020 \\ 
2459558.8285238  &  37.94455\,$\pm$\,0.00298  &  0.17115\,$\pm$\,0.00021 \\ 
2459558.8310499  &  37.94668\,$\pm$\,0.00295  &  0.17107\,$\pm$\,0.00021 \\ 
2459558.8335793  &  37.94689\,$\pm$\,0.00354  &  0.17142\,$\pm$\,0.00023 \\ 
2459558.8360990  &  37.94739\,$\pm$\,0.00346  &  0.17149\,$\pm$\,0.00023 \\ 
2459558.8386285  &  37.94713\,$\pm$\,0.00339  &  0.17097\,$\pm$\,0.00023 \\ 
2459558.8411549  &  37.94462\,$\pm$\,0.00282  &  0.17206\,$\pm$\,0.00020 \\ 
2459558.8436842  &  37.93796\,$\pm$\,0.00270  &  0.17170\,$\pm$\,0.00020 \\ 
2459558.8473803  &  37.93813\,$\pm$\,0.00303  &  0.17137\,$\pm$\,0.00021 \\ 
2459558.8499104  &  37.94440\,$\pm$\,0.00301  &  0.17184\,$\pm$\,0.00021 \\ 
2459558.8524396  &  37.93968\,$\pm$\,0.00292  &  0.17158\,$\pm$\,0.00021 \\ 
2459558.8549660  &  37.93728\,$\pm$\,0.00307  &  0.17101\,$\pm$\,0.00021 \\ 
2459558.8574960  &  37.94566\,$\pm$\,0.00256  &  0.17169\,$\pm$\,0.00019
\enddata
%\tablecomments{placeholder.}
\end{deluxetable}

\begin{deluxetable}{lc}
\tabletypesize{\scriptsize}
\tablewidth{0pt} 
%\tablenum{1}
\tablecaption{Additional RVs of TOI-677 obtained with FEROS. \label{tab:new RVs}}
\tablehead{
\colhead{Time} & \colhead{Radial Velocity}  \vspace{-0.2cm} \\
\colhead{[BJD$_{\textrm{TDB}}$]} & \colhead{[m/s]}
}
%\colnumbers
\startdata 
2459578.84437  &  $38131.6 \pm 12.0$ \\
2459579.83444  &  $38187.8 \pm 12.4$ \\
2459893.84511  &  $38136.6 \pm 12.5$ \\
2459896.82678  &  $37937.8 \pm 14.6$ \\
2459899.84301  &  $37964.6 \pm 10.2$ \\
2459929.79828  &  $37973.1 \pm 10.7$ \\
2459940.79713  &  $38021.7 \pm 10.5$ \\
2459942.86065  &  $37895.6 \pm 9.7$  \\
2459953.77409  &  $37894.0 \pm 10.2$ 
\enddata
%\tablecomments{placeholder.}
\end{deluxetable}

%Appendices


\bibliography{TOI677}{}
\bibliographystyle{aasjournal}



\end{document}